\documentclass[prb,twocolumn,showpacs,preprintnumbers,amsmath,amssymb,superscriptaddress,aps,10pt]{revtex4-1}
\usepackage{graphicx}% Include figure files
\usepackage{dcolumn}% Align table columns on decimal point
\usepackage{bm}% bold math
\usepackage{natbib}
\usepackage{amsmath}
\usepackage{color}
\usepackage{ulem}

\begin{document}
%\preprint{preprint - not for distribution}
%
\title[SERS on CNTs]{Polarized SERS of individual suspended carbon nanotubes by Pt-Re nanoantennas}
\author{Christian B\"auml}
\author{Tobias Korn}
\author{Christoph Lange}
\author{Christian Sch\"uller}
\author{Christoph Strunk}
\author{Nicola Paradiso}\email{nicola.paradiso@physik.uni-regensburg.de}
\affiliation{Institut f\"ur Experimentelle und Angewandte Physik, University of Regensburg}
\begin{abstract}	
We present optical nanoantennas designed for applications that require processing temperatures larger than 800$^{\circ}$C. The antennas consist of arrays of Re/Pt bilayer strips fabricated with a lift-off-free technique on top of etched trenches. Reflectance measurements show a clear plasmonic resonance at approximately 670~nm for light polarized orthogonal to the strip axis.  The functionality of the antennas is demonstrated by growing single-walled carbon nanotubes (CNTs) on top of the antenna arrays and measuring the corresponding Raman signal enhancement of individual CNTs. The results of the measurements are quantitatively discussed in light of numerical simulations which highlight the impact of the substrate.
\end{abstract}
\maketitle
%
%
%
%
%INTRODUCTION
The key feature of plasmonic nanostructures is their ability to confine light on a much smaller scale compared to the far field wavelength\,\cite{Novotnybook,Tame2013}. As a consequence, such structures are able to amplify the optical near field at their surface by many orders of magnitude, especially if a localized surface plasmon resonance (LSPR) is excited. This has a dramatic impact on optical spectroscopy, where the near field amplification makes it possible to detect spectra  of \textit{single molecules}\,\cite{Kneipp1997,Nie1997,Chikkaraddy2016}. In particular, the use of plasmonic structures represented a fundamental breakthrough for Raman spectroscopy, where it is referred to as surface-enhanced Raman spectroscopy (SERS)\,\cite{Moskovits1985,Stosch2011,KneippRoyalSoc}. The name reflects the fact that early experiments made use of granular metallic surfaces\,\cite{Moskovits1985,Fleischmann1974,Creighton1982}. Subsequently, colloidal Au or Ag nanoparticles\,\cite{Creighton1982,Benner1983,Weitz1984}, as well as periodically indented metal layers\,\cite{Sanda1980,Liao1981} were used as plasmonic resonators. Such resonators act as optical nanoantennas, i.e.\,they concentrate propagating radiation into a subwavelength-sized region where the near field dominates the far field.  

The progress of nanoplasmonics enabled the  fabrication of optical nanoantennas specifically designed to work in combination with given target molecules\,\cite{Kneipp1997,Nie1997}.
The combination of plasmonic structures with specific devices has nevertheless a limit: the fabrication of the target device or the synthesis of the desired molecule is not always compatible with the fabrication of the optical nanoantennas.
Carbon nanotubes (CNTs) represent a good example of such a situation. The fabrication of ultra-clean devices\,\cite{[{In the literature CNTs which have never been subjected to any fabrication step and are not interacting with the substrate are called \textit{ultra-clean CNTs}}]SMNote2} often requires that the CNT growth must be performed as last fabrication step\,\cite{Cao2005,Laird2015}. CNTs are grown by chemical vapor deposition (CVD) at temperatures of the order of 800$^{\circ}$C or higher\,\cite{SaitoCNT}. These temperatures are sufficient to melt thin films or nanoparticles of the most common plasmonic materials, namely Au and Ag. Therefore so far the proposed prototypes\,\cite{Kneipp2000,Chu2009,Heeg2014,Paradiso2015} of optical nanoantennas for CNTs had to be fabricated or deposited \textit{after} the CNT growth, i.e.\,they were not compatible with the ultra-clean fabrication scheme.\par 
The low degree of disorder in ultra-clean CNTs allows for a detailed comparison of quantum transport experiments with microscopic theories\,\cite{Laird2015}. For such analysis an independent determination of diameter and chiral angle is highly desirable\,\cite{Cao2004,Dirnaichner2016}. Such information could in principle be provided by Raman spectroscopy, but only in the rare case that the energy of the incident or scattered photon energy matches the energy separation $E_{ii}$ between two van Hove singularities (VHS) of the CNT density of states\,\cite{KatauraIWEPNM,JorioPRL2001,MaultzschPRB2005}.\par 
In experiments with CNTs grown on top of predefined electrodes one deals typically with \textit{few} devices. Without the usage of tunable lasers it is unlikely to meet the Raman resonance condition for an individual contacted CNT. The field amplification induced by plasmonic structures can be crucial to obtain a sizable optical signal from a limited number of contacted devices.\par 
In the present work we developed optical nanoantennas which are resistant to the extreme conditions of the nanotube CVD growth.  We use spatially resolved reflectance measurements to demonstrate the occurrence of a localized surface plasmon resonance (LSPR) at around 630--670~nm. Finally, we exploit the induced field enhancement to %directionally 
magnify the Raman signal of individual ultra-clean CNTs. 
%
%
%
%
%
%
%
%SAMPLE FABRICATION
\section{Experimental details}
\label{sec:expdet}
\begin{figure*}[tb]
\centering
\includegraphics{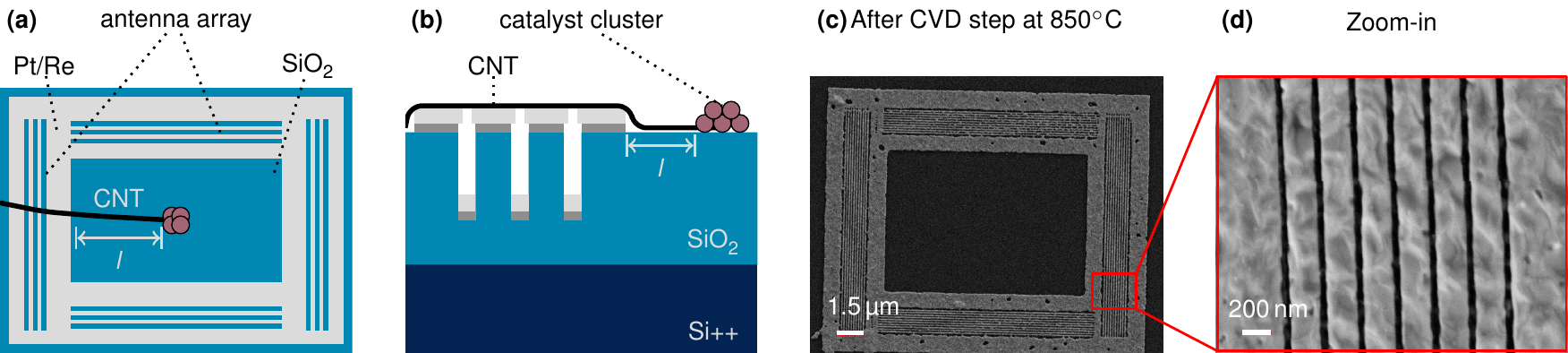}
\caption{(a) Sketch of the sample in top-view. (b) Side-view scheme. Carbon nanotubes (CNTs) start growing 
from clusters of catalyst particles (brown circles) deposited in the middle of the rectangular structure. 
The antenna structures are defined by evaporating a Re-Pt bilayer on top of an array of etched trenches. (c) Scanning electron microscope (SEM) micrograph of a structure after the chemical vapor deposition (CVD) step at 850$^\circ$C. (d) Zoom-in on the right antenna array. The effect of the high temperature of the CVD step is limited to a modest ($\approx5$~nm) corrugation of the metal surface.}
\label{fig:sketch}
\end{figure*}
Our sample fabrication is sketched in Fig.\,\ref{fig:sketch}. The antenna arrays are fabricated on degenerately doped silicon substrates with a 470$\pm 10$~nm-thick SiO$_2$ cap layer. The fabrication process starts with the etching of trenches defined by electron beam lithography (EBL). 30-nm-wide, 100-nm-deep and 6-$\mu$m-long trenches are distributed in arrays of 9~elements with a periodicity of 240~nm. Four arrays are arranged as the four sides of a rectangle.  On top of the trenches, we deposited a Re-Pt (with thickness 7~and 18~nm, respectively) metal bilayer. The choice of Re as sticking layer is motivated by its extraordinary stability at high temperatures. Even when reduced to a few-nanometer-thin layer, Re can stand temperatures as high as 900$^{\circ}$C with negligible structural deformation. On the other hand, at these temperatures Pt nanostructures suffer major roughening, while most metals (including common plasmonic materials as, e.g., Au, Ag, Pd, Al) simply melt. We found that the combination of Re as sticking layer and Pt as top layer is the best compromise between the thermal stability provided by Re, and the plasmonic performances and chemical stability guaranteed by Pt.\par  
Etched trenches make possible to control the gap width between adjacent metal strips: the thicker the metal layer, the thinner is the residual gap\,\cite{Ngoc2013}. The metallization does not cover the entire sample surface, but it is limited to EBL-defined square frames overlapping the trenches, as shown in Fig.\,\ref{fig:sketch}(a,c). In particular the inner part of the rectangular frame is not metallized: here a 1~$\mu$m-wide cluster of catalyst nanoparticles is patterned by EBL\,\cite{[{See the Supplemental Material for further details}]SMnote1}.\par 
The last fabrication step is the CVD growth. This step takes place at 850$^{\circ}$C under a steady stream of CH$_4$ and H$_2$. CNTs start growing vertically from the catalyst cluster. Owing to Van-der-Waals interaction, they bend downwards during the growth until they eventually fall on the substrate or on the antennas\,\cite{Huang2003}.
Most CNTs are shorter than the cluster-to-antenna distance $l\approx 4~\mu$m. However, some CNTs grow long enough to bridge the antennas. Owing to the small gap between the antenna strips, the CNTs are suspended over the substrate. The CNT position and orientation is clearly unpredictable, thus it is determined \textit{a posteriori} by atomic force microscopy (AFM).\par 
Beside the antenna structure described above we have investigated also alternative structures where the nanoantenna arrays are defined by EBL followed by metal deposition and lift-off, i.e.\,without making use of etched trenches. Each strip of the arrays consists a bilayer of 5\,nm of Ti (sticking layer) and 20\,nm of Pt. Although the final results are qualitatively similar, the use of etched trenches makes the fabrication process more reproducible and reliable.  Moreover, etched trenches help to keep the gap between the strips homogeneous and avoid that the roughness caused by the CVD step could short-circuit adjacent strips.

Raman and reflectance measurements are performed in a confocal setup\,\cite{[{See the Supplemental Material for further details}]SMnote1,Stosch2011,Paradiso2015}. 
Light is transmitted through a beam splitter and focused on the sample by a $100\times$ objective lens. The light spot size is approximately $2$\,$\mu$m for white light and below $1\mu$m for the two laser lines, i.e.\,the spatial resolution is comparable to the  array width. The scattered light is transmitted again through the beam splitter and then sent to the detector, as discussed in our previous work\,\cite{Paradiso2015}. 
The configuration of light sources, detectors and polarizers depends on the specific measurement\,\cite{[{See the Supplemental Material for further details}]SMnote1}. For reflectance spectroscopy we use a thermal source of white light and a spectrometer as detector. An analyzer is placed in between the sample and the spectrometer. For reflectance maps the light source is a laser (either a He/Ne laser with $\lambda_L=633$\,nm or a diode laser with $\lambda_L=532$\,nm) and the detector is a power--meter. A polarizer is placed between the source and the sample. The same configuration is used for Raman spectroscopy. In this case the detector is a spectrometer.

\section{Experimental results}
\begin{figure}[tb]
\includegraphics{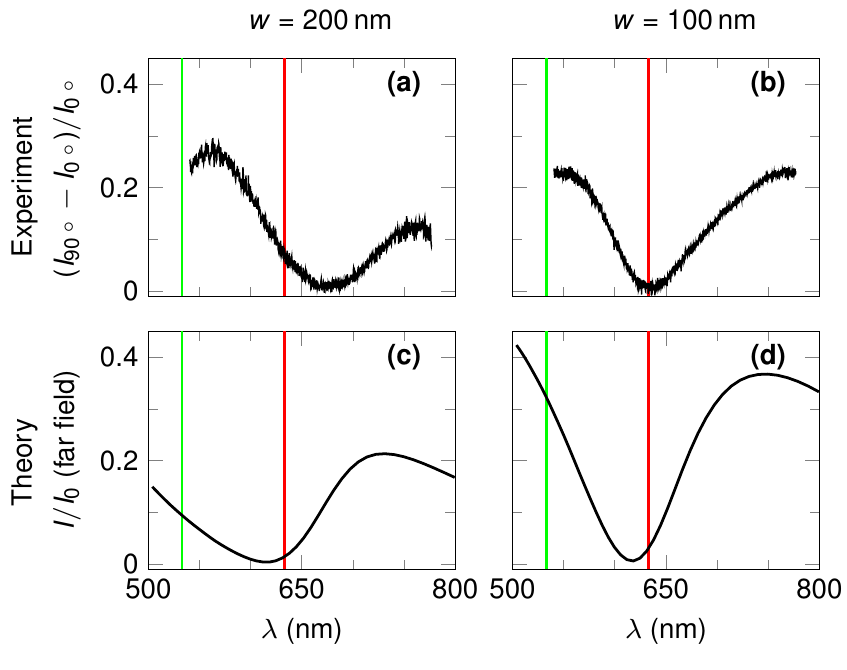}
\caption{We illuminated a vertical and a horizontal antenna array by unpolarized white light and we measured for both structures the spectrum of the  vertical polarization component by means of an analyzer. Panels (a) and (b) show the difference between the signal scattered by the horizontal array ($I_{90^{\circ}}$) and the vertical array ($I_{0^{\circ}}$), which serves as background. The curves are normalized to $I_{0^{\circ}}$. The graph in (a) refers to a Re-Pt antenna array whose width is $w=200$~nm. The graph in panel  (b) refers to an array of 100~nm-wide antennas defined by EBL. Graphs in (c) and (d) show the corresponding results of numerical simulations (see Fig.\,\ref{fig:simul}), which are discussed in Sec.\,\ref{sec:discussion}. The colored vertical lines indicate the laser wavelengths 532\,nm (green) and 633\,nm (red).}
\label{fig:white}
\end{figure}
\begin{figure*}[tb]
\includegraphics{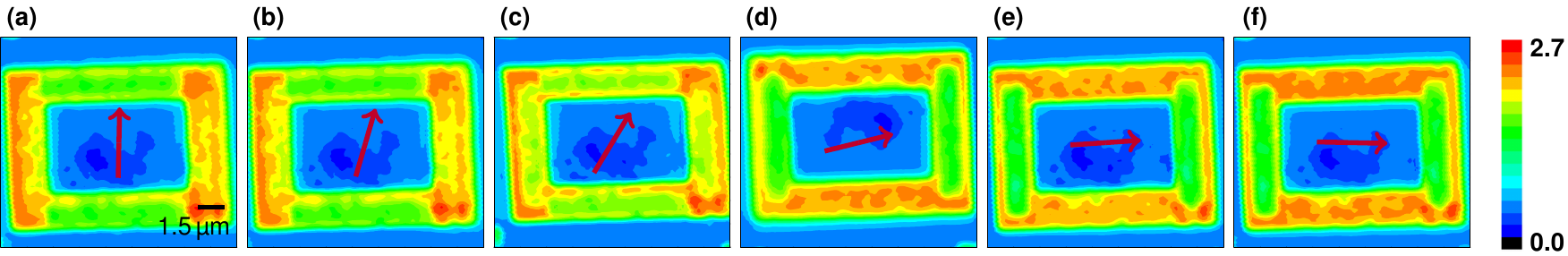}
\caption{Reflectance maps for incident wavelength $\lambda_L=633$~nm and for different polarization directions, indicated by the red arrow. The maps have been acquired by scanning the laser spot over the sample with steps of 200~nm. The reflected signal is maximal for polarization direction parallel to the array axis and minimal for polarization perpendicular to the array axis. This effect is  not observed for $\lambda_L=532$~nm, where no plasmonic resonance is excited\,\cite{[{See the Appendix for further details}]SMnote1}. The reflectance is normalized to that of the substrate.}
\label{fig:maps}
\end{figure*}
Long metal nanowires display a plasmonic resonance near the visible when excited with a field orthogonal to their axis, where the incompressible electron plasma experiences a non-negligible restoring force\,\cite{Novotnybook,Ngoc2013}. The excitation of a LSPR implies losses due to absorption in the metal, which results in a  reduction of the scattered light intensity. If a metal nanowire is illuminated with white light polarized orthogonally to the wire axis, plasmonic resonances will appear therefore as minima in the spectrum of the reflected light. On the other hand, when an antenna array is excited with unpolarized light, only the polarization component perpendicular to the array axis will show a minimum at the LSPR, while the parallel component will be scarcely affected.\par  

In our reflectance spectroscopy measurements  we keep the analyzer oriented vertically [i.e.\,parallel to the shorter side of the rectangle in Fig.\,\ref{fig:sketch}(c)] and we measure the reflected signal from two arrays, one oriented horizontally ($I_{90^{\circ}}$) and the other oriented vertically ($I_{0^{\circ}}$), where the latter represents the background signal. The difference of the two signals  is then normalized to the background signal ($I_{0^{\circ}}$)\,\cite{[{The same result can be obtained by rotating the analyzer on the \textit{same} array. The method here described is however more accurate since the optical transmission of our setup is slightly polarization and wavelength dependent}]SMNote3}.
Figure\,\ref{fig:white}(a,b) shows the difference between $I_{90^{\circ}}$ and $I_{0^{\circ}}$, divided by the latter. The panel (a) refers to a Re/Pt antenna array patterned on etched trenches. The graph clearly displays a minimum in the reflected signal at $\approx 670$~nm, which is close to the laser wavelength $\lambda_L = 633$~nm used in our Raman experiments. The graph in Fig.\,\ref{fig:white}(b) shows data obtained on a EBL-defined antenna array, i.e.\,without etched trenches. In this case the minimum occurs at a slightly lower wavelength, $\approx 630$~nm. Graphs in Fig.\,\ref{fig:white}(c) and (d) show the corresponding results of the numerical simulations discussed in Sec.\,\ref{sec:discussion}.

To confirm the resonant  nature of the observed minimum, we acquired maps of the reflected signal as a function of the polarization of the incident light. In this case we use monochromatic light from a laser source (with $\lambda_L=633$ or 532~nm), focused to a 
 0.5~$\mu$m-wide spot. Figure\,\ref{fig:maps} shows how reflectance maps evolve when the  polarization of the incident light at $\lambda_L=633$~nm is gradually rotated from vertical (a) to horizontal (f). The maps clearly show that horizontal antennas have a minimum in the reflected signal for incident light polarization directed vertically, and vice versa. This corresponds to the excitation of a LSPR as discussed in the previous paragraph. By rotating the incident light polarization the reflected signal from the horizontal antennas continuously increases, while the signal from the vertical ones decreases accordingly. This behavior is not observed for the green laser ($\lambda_L=532$~nm), because this wavelength is far from the LSPR minimum\,\cite{[{See the Appendix for further details}]SMnote1}.\par 
%
%
%
%
%RAMAN MEASUREMENTS
%
%
\begin{figure}[tb]
\includegraphics{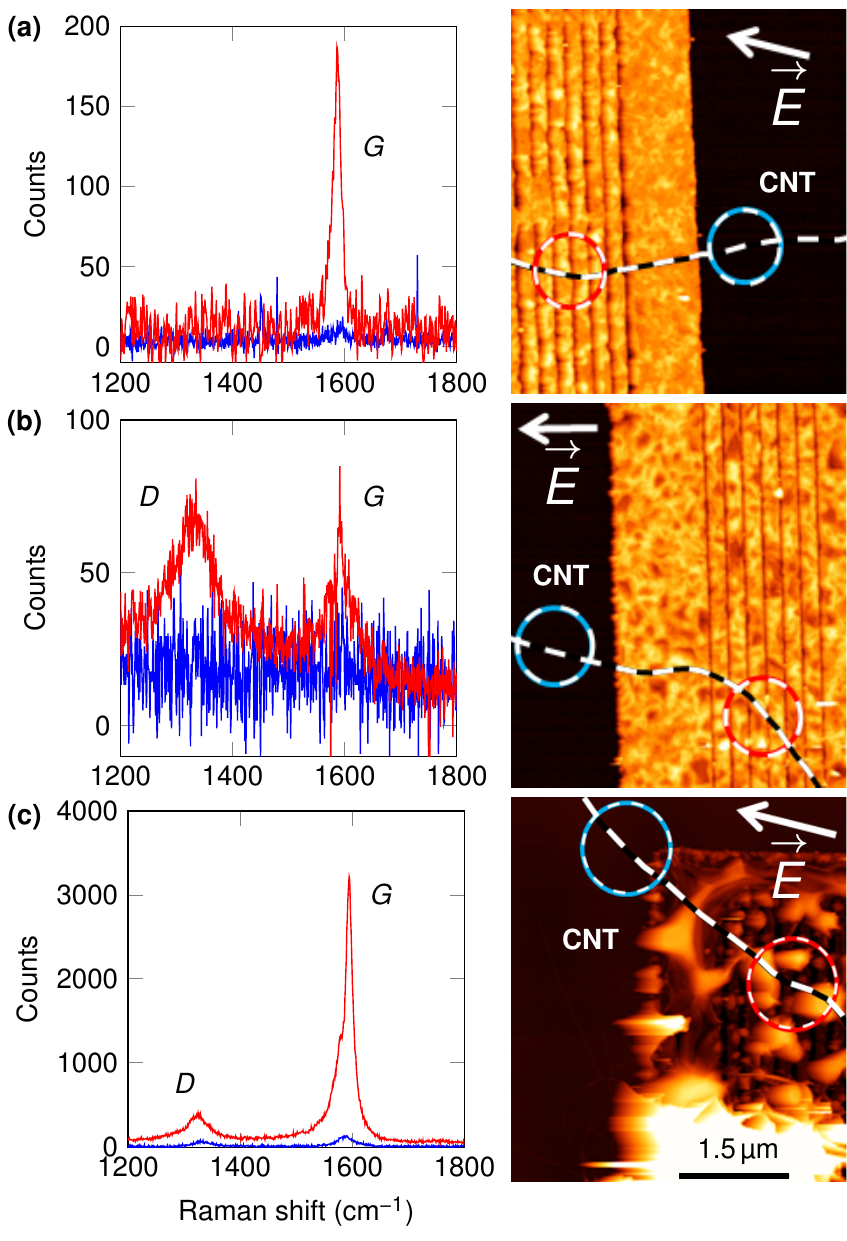}
\caption{Raman spectra for CNTs overgrown on optical antenna arrays (left panels) and corresponding AFM topography images (right panels). 
Red and blue curves refer to measurements on CNT portions overgrown on antennas or lying on the substrate, respectively.   The red and blue circles indicate the position and the size of the focused light spot for the corresponding measurement. The white arrows indicate the polarization directions. In (a) and (b) the antenna strips consist of a Re/Pt bilayer. In (c)  antenna strips consist of Ti/Pt  and are defined by EBL without etched trenches. Compared to the structures above, the corrugations due to the CVD process are much more pronounced. All the Raman spectra have been obtained after 30~s of integration, except the bare CNT signal in panel (a), where the time was 10 times longer and the signal was then divided by a factor 10.}
\label{fig:raman}
\end{figure}
Having demonstrated the occurrence of a plasmonic resonance, we used antenna arrays to amplify the Raman signal of selected suspended CNTs. After the CVD growth process, AFM scans are used to locate the presence of individual CNTs and where exactly they cross the antenna arrays. A convenient CNT is in this case one far from other CNTs and long enough completely bridge the antenna array.\par  
Figure\,\ref{fig:raman} shows the results of Raman measurements performed on three samples. The red curves on the left panels refer to the Raman signal around the $G$ and the $D$ peak measured on CNT segments located on the antenna strips, while the blue ones correspond to Raman measurements on bare portions. The spectra have been acquired after an integration time of 30\,s. The right panels in Fig.\,\ref{fig:raman} show AFM topography scans of the three samples, where the red and light blue circles indicate the laser spot size and location for the bare CNT portion and for the CNT on antennas, respectively. The arrows indicate the  incident light polarization for each measurement. The difference between red and blue curves in Fig.\,\ref{fig:raman}(a)--(c) clearly shows the dramatic amplification of the Raman signal induced by the optical antennas.\par 
In order to quantify the signal amplification, we define the enhancement factor as the ratio between the Raman peak amplitudes measured with and without antennas. 
The enhancement factor is difficult to estimate when the signal on the bare portion of the CNT is too weak. The intensity of the Raman signal measured on a bare CNT depends on the difference between the incident $\hbar \omega_L$ (or the scattered $\hbar \omega_{L} \pm \hbar\omega_{q}$) photon energy and the energy separation $E_{ii}$ between VHSs. The width of such resonance windows is roughly of the order of 200~meV for the $G$ mode and around 100~meV for the radial breathing mode (RBM)\,\cite{MaultzschPRB2005}.
If the difference between incident (or scattered) photon and the transition energy $E_{ii}$ is larger than the above values, then the Raman signal is drastically suppressed. This is often the case: for a given  laser frequency $\omega_{L}$, only a small fraction of the CNTs shows a significant signal. This is precisely where the advantage of optical antennas becomes obvious, as the signal amplification compensates for the suppression due to the energy mismatch~\cite{Paradiso2015}.  As demonstrated in Fig.\,\ref{fig:raman}, also CNTs which are scarcely or not measurable without antennas provide a significant signal. From the spectrum in  Fig.~\ref{fig:raman}(a) we deduce an enhancement factor of around 40. While the signal on the bare portion of this CNT is barely discernible, the amplified signal provides useful information: the CNT has a low number of defects, as deduced by the absence of the $D$ peak; furthermore the narrow (FWHM:~17 cm$^{-1}$) $G$ peak implies a very weak $G^{-}$ components, which indicates either an achiral CNT (armchair or zig-zig) or a CNT with very low chiral angle\,\cite{Park2009}.\par 
The graph in Fig.\,\ref{fig:raman}(b) refers to a CNT whose Raman spectrum is not measurable on the bare portion even after an integration time of 300\,s, i.e.\,ten times longer than the usual one. This indicates that the CNT has no $E_{ii}$ transitions near 633~nm. However, the signal on the CNT portion suspended on the antennas is clearly detectable, though weak. Since the $D$ peak is stronger than the $G$ peak, we deduce that this CNT has a relatively high number of defects.\par  
Finally Fig.\,\ref{fig:raman}(c) shows the results of Raman measurements on antennas made of Ti/Pt, i.e.\,without the Re sticking layer. As mentioned above, the absence of the Re sticking layer induce large corrugations in the Pt film when subjected to growth temperatures of the order of 850$^{\circ}$C. We note that despite the surface roughness the enhancement factor remains considerably high, of the order of 60.  This particular CNT has a separation $E_{ii}$ between VHSs relatively close to the incident photon energy, thus the Raman signal is large enough to be detected without antennas as well. However, the antenna array allows us to measure also the RBM (not shown). By following the standard assignment procedure we deduce the possible chiral indices for the CNT. We found that the most likely chiral indices are (22,5) or (21,7), two adjacent elements of the CNT family $2n+m=49$ which share very similar properties, as e.g.~diameter, chiral angle and energy separation between VHSs\,\cite{[{See the Supplemental Material for further details}]SMnote1}. 
\section{Discussion}
\label{sec:discussion}
\begin{figure*}[tb]
\includegraphics{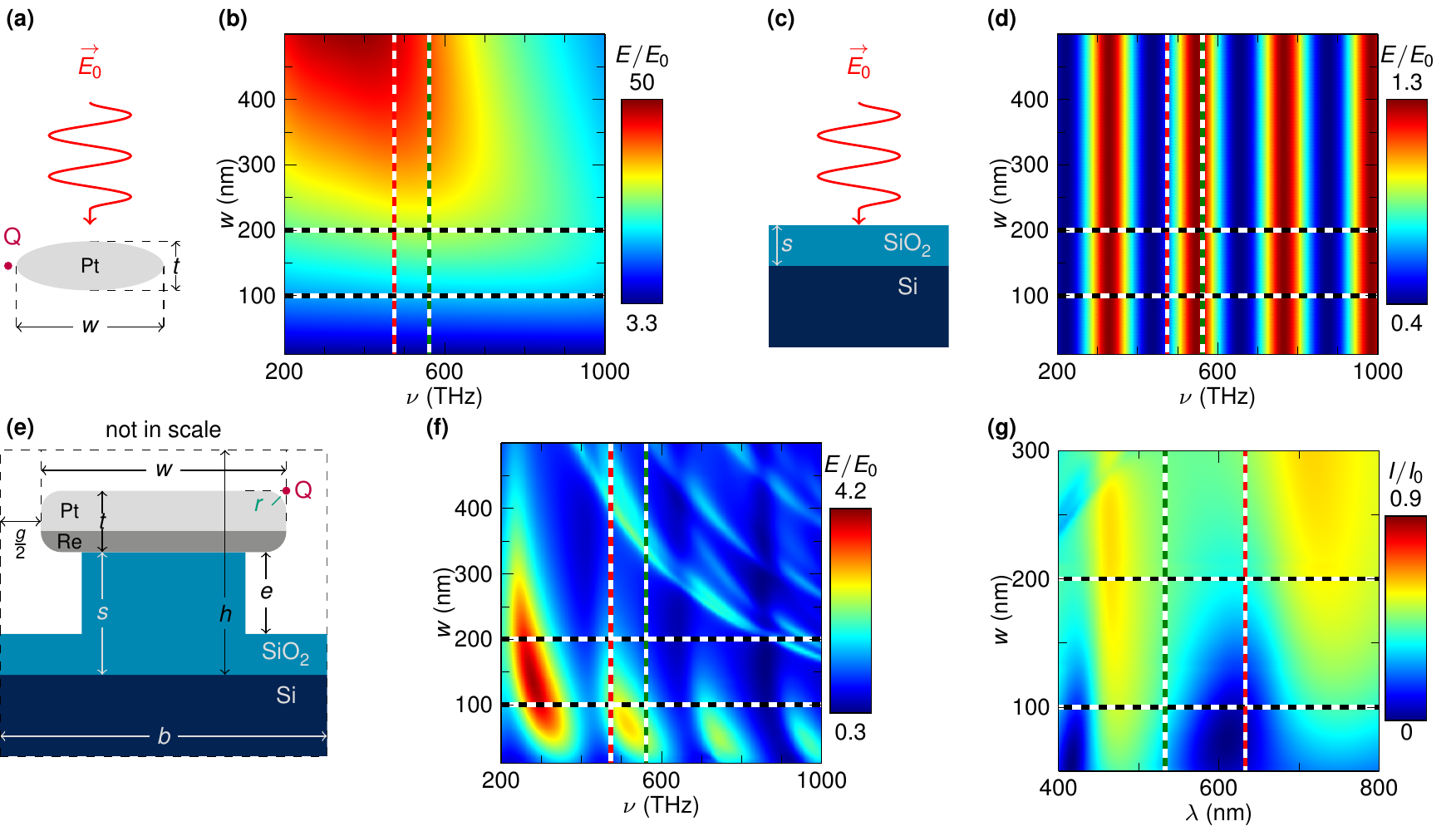}
\caption{
(a) Sketch of the geometry considered in the analytical calculations plotted in panel (b). An infinitely long Pt strip of elliptical section is illuminated from above by a plane wave of amplitude $|E_0|$. The incident field polarization is perpendicular to the strip axis. The vertical axis $t=25$\,nm equals the thickness of the strips in our experiment. The point Q indicates the position where the field strength is calculated. (b) Corresponding color plot of the electric field intensity calculated in the point Q at 5~nm from the strip edge.  The field is plotted as a function of strip width $w$ and excitation frequency $\nu$.   
(c) In a complementary calculation we consider the effect of the substrate alone (see text). The substrate consists of a doped Si wafer capped with a 470\,nm-thick SiO$_2$ layer. (d) The corresponding color plot for the electric field amplitude at the top surface of the SiO$_2$ layer. 
(e) Sketch of the geometry used in our finite-difference frequency-domain (FDFD) simulation. We used the following parameters: gap $g=40$~nm, etching depth $e=100$~nm, SiO$_2$ layer thickness $s=470$~nm, strip thickness $t=25$~nm (the Re and Pt layers are 7 and 18~nm thick, respectively) The simulation box width and height are $b=240$~nm and $h=1000$~nm, respectively. The corners of the metal  strip have been rounded (curvature radius $r=5$~nm). The width $w$ is swept from 10 to 500~nm and the field frequency $\nu$ is swept from 200 to 1000\,THz.
(f) Electric field amplitude calculated by FDFD in the point Q located along the strip axis at 5~nm from the metal edge. 
(g) Result of the FDFD simulation for the reflected \textit{far field} intensity $I$ normalized to the incident intensity $I_0$ plotted as a function of $w$ and wavelength $\lambda$. In all plots the frequencies (or wavelengths) of our lasers are indicated with corresponding colored dashed lines.  Two black horizontal dashed lines indicate antenna widths of 100 and 200~nm respectively.
}
\label{fig:simul}
\end{figure*}

The widths of the plasmonic resonances measured in our reflectance spectroscopy experiments are of the order of $\Delta \lambda\approx 100$~nm.  
As shown below, simple model calculations considering metallic strips alone lead to plasmonic resonance widths much larger than the observed ones. One possibility to explain the observed narrow resonances is to consider in addition  the optical interference due to the SiO$_2$/Si interface 470\,nm below the sample surface. We will show that the interplay between the optical mode in the SiO$_2$ cavity and the plasmonic mode within the antenna strips gives rise to much sharper eigenmodes. Numerical simulations, discussed at the end of this section, show that such modes have a width comparable to that observed in the experiment. 

In our analytical calculation\,\cite{[{See the Supplemental Material for further details}]SMnote1} we approximate a single strip as an infinitely long cylinder of elliptical cross-section. 
As sketched in Fig.\,\ref{fig:simul}(a) the field is evaluated in the position Q located 5~nm away from the strip edge. The vertical axis $t$ measures 25~nm (equal to our antenna thickness) and the horizontal axis $w$ is varied. The graph in Fig.\,\ref{fig:simul}(b) shows the results of the calculation for the electric field amplitude (normalized to the incident field amplitude) as a function of both $w$ and the incident frequency $\nu$. We notice that, owing to the large imaginary part of the refractive index of Pt\,\cite{Rakic98}, for all widths $w$ the LSPR resonance width is much broader than that observed in our experiment. 

In principle, the plasmonic eigenmodes of an array of strips differ from those of a single structure owing to the electrostatic interaction between the strips which leads to mode hybridization. However, such a difference is relevant only if the decay length for the near field amplitude in the vicinity of a strip is comparable to the gap between two adjacent strips. In the visible range such decay length for submicrometric plasmonic structures is of the order of a few nanometers\,\cite{[{See the Supplemental Material for further details}]SMnote1,Novotnybook}, which is much smaller than the gap ($g=40$~nm). Therefore the interaction between  adjacent strips can be neglected in our case.\par 

The sharp minima observed in our reflectance spectroscopy experiment can be explained by considering the effect of the substrate. Indeed, even in a plain SiO$_2$/Si substrate the electric field amplitude at the sample surface strongly depends on the frequency $\nu$, owing to the optical interference between the two surfaces of the SiO$_2$ film. In Fig.\,\ref{fig:simul}(d) we plot the calculated electric field amplitude in a point located on the upper SiO$_2$ surface. For ease of comparison with one of the following figures, the graph is plotted  as a function of $w$ and $\nu$ as in Fig.\,\ref{fig:simul}(b), although in absence of nanoantennas the variable $w$ plays no role. The graph shows that the field amplitude at the surface considerably oscillates as a function of the frequency. Clearly, this significantly alters the LSPR profile of optical antennas patterned on top of a SiO$_2$ film.\par  
To quantify the impact of the substrate on the LSPR, we performed a finite-difference frequency-domain (FDFD) numerical simulation on the actual geometry, sketched in Fig.\,\ref{fig:simul}(e). 
In the calculation we model the nanoantennas as periodic arrays of infinitely long strips. By sweeping geometry parameters and laser frequency, we do not only extract field enhancement factors but also identify plasmonic modes and the interplay between the metal structure and the SiO$_2$/Si substrate stack. 
The strip section is assumed to be rectangular with rounded corners (with radius of curvature $r=5$~nm). The gap $g=40$~nm, the etching depth $e=100$~nm and the SiO$_2$ thickness $s=470$~nm are kept constant. The result of the simulation is plotted in Fig.\,\ref{fig:simul}(f). The graph shows the horizontal component of the electric field amplitude calculated in the point Q indicated in the sketch, plotted as a function of  width $w$ and frequency $\nu$. A comparison with the graphs in Fig.\,\ref{fig:simul}(b) and (d) reveals that the actual eigenmodes are non-trivial combinations of the cavity modes in the SiO$_2$ layer and the plasmonic modes in the metal strip. In fact, the graph in Fig.\,\ref{fig:simul}(f) shows a broad feature (on the left and bottom part of the graph) modulated by almost vertical fringes which are clearly related to the SiO$_2$ film interference.   As a result, in the visible range the maximum of the near field amplitude for a 200~nm-wide strip occurs at approximately  470~THz, i.e.\,close to the red laser frequency. We also notice several arc-shaped features on the top-right zone of the graph, which correspond to higher energy-modes. 
Further details about the calculation and the interpretation of optical modes  are given in the Supplemental Material.

The reflection coefficient for a thin film deposited on top of a high refractive index substrate displays a maximum when the electric field at the surface is minimal\,\cite{[{See the Supplemental Material for further details}]SMnote1}. This condition corresponds to a frequency $\nu$ such that the electric field forms a standing wave within the SiO$_2$ layer, with nodes at the two interfaces. Vice versa, when the reflection coefficient is minimal, the electric field at the surface is maximal. On the other hand, when the polarizability $\alpha$ of an optical antenna (and thus the near field amplitude) has a maximum as a function of $\nu$ then, due to the imaginary part of the metal refractive index, the reflected far field shows a minimum. From these arguments we expect that the color plot of the \textit{reflected far field}, shown in Fig.\,\ref{fig:simul}(g), will display a reverse contrast when compared to that of the near field shown in Fig.\,\ref{fig:simul}(f). In order to compare the simulation results with the white light reflectance spectra, in Fig.\,\ref{fig:simul}(g)  we plot the intensity as a function of the \textit{wavelength} $\lambda$ and strip width $w$. The line cuts for $w=200$ and 100~nm are shown in  Fig.\,\ref{fig:white}(c,d). These latter indicate that the minimum for the reflected signal occurs at $\approx 620$~nm, the resonance width is approximately 100~nm and that the curve for  $w=100$~nm is slightly sharper and slightly blue-shifted compared to the one for  $w=200$~nm. These features are in agreement with the experimental data. We notice, however, that the precise shape of the minimum is not suitably captured by the simulation. Note that the exact wavelength $\lambda_m$ for the minimum critically depends on the SiO$_2$ layer thickness. The SiO$_2$ layer thickness is measured with an accuracy of 10~nm, which causes a comparable uncertainty for $\lambda_m$ in the simulation.\par 
A conclusion drawn from the calculations above is that the dependence of the antenna resonance frequency on both  antenna width and antenna gap is relatively smooth. Therefore the roughness produced by the CVD process has a negligible effect on both the resonance frequency and the enhancement factor. This can explain the good enhancement factor observed in Fig.\,\ref{fig:raman}(c) despite the surface roughness of the metal stripes of that sample.
%
%
%
%
%
%
%
%
%CONLUSIONS
\section{Conclusions}
We have demonstrated directional optical antennas for applications that require extremely high process temperatures as,  e.g., those required for the CVD growth of CNTs. We have fabricated devices where CNTs are grown on top of antenna arrays and shown that the latter significantly amplify the CNT Raman signal. Numerical simulations show that the relatively sharp antenna resonance is due to the interplay of plasmonic resonance and thin film interference  due to the SiO$_2$ cap layer.
Possible applications of Pt-Re optical antennas go well beyond SERS of CNTs, since the present fabrication scheme can be in principle applied to any nanostructure for optical spectroscopy whose fabrication requires extreme temperature conditions.
%
%
%
%
%ACKNOWLEDGEMENTS
\begin{acknowledgments}
The work was
funded by the European Union within the STREP “SE2ND”,
and by the Deutsche Forschungsgemeinschaft within Grants
No.\,SFB689 and No.\,GRK1570.
\end{acknowledgments}

\bibliography{biblio}

\end{document}